\documentstyle[prl,aps,epsf,multicol]{revtex}
\def\beginwide{ \end{multicols} \widetext\noindent}
\def\endwide{ \begin{multicols}{2} \narrowtext\noindent }
\font\smallfont=cmr9
\begin{document}
\title{
 Solution of quantum double exchange on the complete graph from spl(2,1) dynamical supersymmetry}
\author{Karlo Penc$^1$\cite{*} and Robert Lacaze$^{1,2}$}
\address{$^1$ Service de Physique Th\'eorique, CEA-Saclay, 91191 Gif-sur-Yvette Cedex, France}
\address{$^2$ASCI, Bat. 506, Universit\'e Paris Sud, 91405 Orsay Cedex, France}
\date{30 May 1998}
\maketitle
\begin{abstract}
The exact and analytical low energy spectrum  of  
the ferromagnetic Kondo lattice model on a complete graph extended 
with on-site repulsion $U$ is obtained in terms of a dynamical 
spl(2,1) supersymmetry in the limit of infinitely 
strong Hund's coupling (double exchange). Furthermore, we show that for the particular value
of $U=J_H/2$ the supersymmetry is not constrained to infinite $J_H/t$ only and we
calculate the energy including the $t^2/J_H$ corrections analytically 
and we give numerical evidence which suggest that the Kondo Hamiltonian is
itself supersymmetric for any $J_H/t$. On a $N$ site graph, the ferromagnetic ground state is realized for 1 and $N$+1 electrons only.
In the leading order in the value of the core spin, the quantum and semiclassical spectra are identical.
\end{abstract}

{\smallfont PACS Nos : 71.27.+a,75.30.Et,75.30.Mb,03.65.Fd \hfill
T98/042 \quad  Submitted to Phys. Rev. Letters}

\widetext
\begin{multicols}{2}

\narrowtext
The recent  experimental activity on manganites, La$_{1-x}A_x$ MnO$_3$
(where $A$=Ca,Sr or Ba), a family intensively studied due to its colossal
magnetoresistance\cite{exp}, has stimulated the interest of theorists for
this compound.  While the orbital degrees of freedom certainly cannot be 
neglected for the real material\cite{millis}, it is believed that some
aspects of the physics of the transition metal oxides can be 
revealed by considering the Kondo-lattice Hamiltonian
${\cal H}_{\rm KL} = {\cal T} + {\cal H}_{\rm int}$,
where the kinetic part
\begin{equation}
  {\cal T} = -\sum_{i,j,\alpha} t_{ij} 
    c^{\dagger}_{i\alpha} c^{\phantom{\dagger}}_{j\alpha}, \qquad t_{ij}>0,
  \label{eq:HKLkin}
\end{equation}
describes the hopping of electron on a lattice 
($\alpha=\uparrow,\downarrow$ is the spin index), and the interaction part reads
\begin{equation}
  {\cal H}_{\rm int} =
 -\frac{J_H}{2} \sum_{j,\alpha,\beta}  
  c^{\dagger}_{j\alpha} 
   {\bf S}_{cj} \bbox{\sigma}^{\phantom{\dagger}}_{\alpha\beta} 
   c^{\phantom{\dagger}}_{j\beta}
 + U \sum_{j} n_{j\uparrow} n_{j\downarrow} .
  \label{eq:HKLint}
\end{equation}
Here $J_H$ stands for the intratomic ferromagnetic exchange (Hunds coupling) \
between the conduction electrons and localized core electrons\cite{hund},
$\bbox{\sigma}_{\alpha\beta}$ denotes the vector of
Pauli matrices, and ${\bf S}_{cj}$ is the spin operator of the localized 
core electrons with spin $S_c$. 
We also include the on-site Coulomb repulsion $U>0$ between the 
electrons, with $U$ of the order of $J_H$.
While here we are interested in  $J_H>0$, the Kondo-lattice Hamiltonian 
has been extensively studied for
$J_H<0$, as describing heavy fermion systems\cite{revKL}. 

Because the Hund coupling is typically larger than the  hopping
(in manganites $J_H\approx 1 - 2$eV
and $t\approx 0.1-0.5$ eV), the energetically unfavorable low spin 
states are usually neglected and we get the quantum double exchange, with a rather
complicated Hamiltonian \cite{kkubo,mullerhartmann}. The next usually used approximation
neglects the quantum spin fluctuation of the $S_c$ core spin 
replacing it by classical variables
(spherical angles $\theta$ and $\phi$). These two approximations lead to 
the double-exchange model \cite{mullerhartmann,doubleexch} with Hamiltonian 
$H_{\rm de}=-\sum \tilde t_{ij} f^\dagger_i f^{\phantom{\dagger}}_j$ , 
which describes noninteracting spinless fermions (charges) 
moving in a disordered background of classical spins
$ \tilde t_{ij} =  t_{ij} 
   \left[ \cos(\theta_i/2)\cos(\theta_j/2)+
 \sin(\theta_i/2)\sin(\theta_j/2)e^{i(\phi_i-\phi_j)}
              \right].$
The charges can freely propagate provided the core spins are aligned and 
therefore  ferromagnetism is favored. 
The main effect of finite $J_H$ is to introduce 
antiferromagnetic exchange between the core spins, which will hinder the 
free propagation of the holes, resulting in a competition between 
ferromagnetic and antiferromagnetic ordering. 
 To check the scenario presented above, different numerical methods, like
 exact diagonalization, quantum Monte Carlo, density matrix renormalisation 
group and dynamical mean field are applied
\cite{dagotto,furukawa,karen,horsch}.

\begin{figure}[t]
 \epsfxsize=8.5 truecm
 \centerline{\epsffile{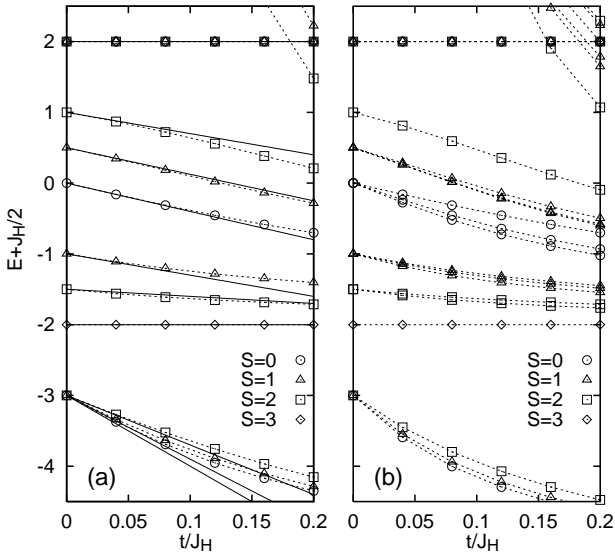}}
 \caption{The low energy spectrum of the ferromagnetic Kondo lattice on a 
4 site complete graph with $S_c=1/2$ and $N_e=2$
for $U=J_H/2$ (a) and $U=0$ (b). 
Each point is highly degenerate.
The solid straight lines show the energy of the effective Hamiltonian as given by Eqs. (\ref{eq:0nes})-(\ref{eq:enes}).}
 \label{fig:1}
\end{figure}

Here we will consider the model where the 
hopping is infinitely long ranged, i.e. complete graph with $t_{ij}=t(1-\delta_{ij})$.
The numerical diagonalizations show that the spectrum separate into well defined
bands for large enough $J_H$. 
A typical example of the lowest band is shown in Fig.~\ref{fig:1} in the case
of a small 4-site cluster
where states at the right upper corner belong to the next band.
It appears that with long ranged hopping, the lowest band is surprisingly simple
in the limit $J_H/t \to \infty$, and $U$ independent with huge
degeneracies.
 The distribution of the eigenvalues is
analogous to the spectrum of the $U/t \to \infty$ Hubbard model ($t$-model)
on a complete graph \cite{kirson,tmodel} and, as shown later, this
peculiar spectrum reflects the same underlying spl(2,1) dynamical supersymmetry
\cite{kirson} (supersymmetry in the sense of odd and even number of fermions
mixed in the same irreducible representation).
For finite $J_H/t$ the degeneracies are lifted except in the
special case $U=J_H/2$ where some states with the same total spin
value $S$ remain degenerated as it can be realized in the qualitative comparison of Figs
1a and 1b.
To understand this 
behavior: (a) we introduce Schwinger bosons to describe the 
spin degrees of freedom, and the form of the effective strong coupling Hamiltonian 
becomes remarkably simple, especially for $U=J_H/2$;
 (b) the effective strong coupling Hamiltonian is then solved  exactly using 
dynamical supersymmetry and the spl(2,1) graded algebra; (c) we show that in
the limit $S_c \to \infty$ the semiclassical and quantum model give the 
same energy
\noindent
spectrum, and that the semiclassical limit can be
performed already at the level of the Kondo lattice Hamiltonian.
\paragraph{Effective strong coupling Hamiltonian.}
 In the atomic limit ($t=0$), the different lattice 
sites decouple.
An empty site has energy $0$; one electron can form with the core spin
either the high ($S_c+1/2$) or low ($S_c-1/2$) spin state, with $E=- J_H S_c/2$ and 
$E=  J_H (S_c+1)/2$, respectively; finally two electrons 
results in a state with energy $E= U$ and spin $S_c$. 
This can be summarized by representing the different states using 
auxiliary fermions $f$ and $d$ and the spins by Schwinger bosons $b_{\sigma}$\cite{sarker}:
\begin{eqnarray}
 c^{\dagger}_{j\uparrow} &=& \frac{1}{\sqrt{2 S_c+1}} 
   \left( b^\dagger_{j\uparrow} f^\dagger_j 
        + b^{\phantom{}}_{j\downarrow} d^\dagger_j \right), \nonumber\\
 c^{\dagger}_{j\downarrow} &=& \frac{1}{\sqrt{2 S_c+1}} 
   \left( b^\dagger_{j\downarrow} f^\dagger_{j}
          - b^{\phantom{}}_{j\uparrow} d^\dagger_j \right). 
  \nonumber
\end{eqnarray}
where $f^\dagger$ and $d^\dagger$ are the creation operators of high and low spin
states respectively while the spin operators in terms of Schwinger bosons are
$S^z_j = (n^b_{j\uparrow}-n^b_{j\downarrow})/2$, 
$S^+_j = b^\dagger_{j\uparrow}b^{\phantom{}}_{j\downarrow}$,
and $S^-_j = b^\dagger_{j\downarrow}b^{\phantom{}}_{j\uparrow}$.
The anticommutation relation for electrons require the following constraint
to be satisfied at each site
\[
  \sum_\alpha n^b_{j\alpha} - n^f_j + n^d_j = 2 S_c . 
\]
In this representation the  interaction part becomes diagonal
\[
  {\cal H}_{\rm int} =  \frac{J_H}{2} 
  \sum_j\left[(S_c+1) n^d_j - S_c n^f_j - n^f_j n^d_j \right]  
   +U \sum_j n^f_j n^d_j .
\]
Choosing $U=J_H/2$ we can eliminate the  four fermion term $n^f_j n^d_j$.
The effective strong 
coupling Hamiltonian is then the expansion in $t/J_H$ around the atomic 
limit and is obtained using a canonical transformation. 
As a first step, following Ref.{\onlinecite{oles94}},  we decompose 
the kinetic Hamiltonian into a part conserving the number of $d$ and $f$ 
fermions (${\cal T}_0$),
and a part where the final and initial state differs by a large 
energy $J_H (S_c+1/2)$ 
\[
  {\cal T}_{J} = -\sum_{ij} \frac{t_{ij}}{2S_c+1}
   \left( b^{\phantom{\dagger}}_{i\downarrow} b^{\phantom{\dagger}}_{j\uparrow} - 
          b^{\phantom{\dagger}}_{i\uparrow} b^{\phantom{\dagger}}_{j\downarrow} \right)
    d^\dagger_i f^{\phantom{\dagger}}_j  \ ,
\]
 so that 
$  {\cal T} = {\cal T}_0 + {\cal T}_{J} + {\cal T}_{J}^\dagger $.
 The effective Hamiltonian is then given by 
${\cal H}_{\rm eff} ={\cal H}_{\rm int} + {\cal T}_0  - 
     \left[ 
       {\cal T}^\dagger_{J},
       {\cal T}_{J}
     \right] /[(S_c+1/2)J_H] + {\cal O}(t^3/J_H^2)$, and 
in the lowest energy subspace, where we keep the $f$ 
fermions only, it reads
\beginwide
\begin{equation}
  {\cal H}_{\rm eff} = - \frac{J_H N_e S_c}{2}
    -\frac{1}{2 S_c\!+\!1} 
  \sum_{i,j,\alpha} t_{ij} 
     f^\dagger_i b^\dagger_{i\alpha} 
     b^{\phantom{\dagger}}_{j\alpha}  f^{\phantom{\dagger}}_j
     -  \frac{2}{J_H (2S_c\!+\!1)^3} \sum_{i,j,k,\alpha,\beta}
 \! \! \! t_{ki} t_{ij} 
   f^\dagger_{k} b^\dagger_{k\alpha} 
   \left[
      \left(S_c \! + \! \frac{n^f_i}{2}\right)\delta_{\alpha\beta}
       -{\bf S}_i \bbox{\sigma}_{\alpha\beta}    \right] 
   b^{\phantom{\dagger}}_{j\beta} f^{\phantom{\dagger}}_j
 \label{eq:Heff}
\end{equation}
\endwide
The first order term proportional to $t$ in the expansion is equivalent to
quantum double-exchange Hamiltonian  
derived by Kubo and Ohata \cite{kkubo} and by M\"uller-Hartmann and 
Dagotto\cite{mullerhartmann}. Let us note here, that the procedure can be 
repeated for general $U$, and it would give us a more complicated 
$t^2/J_H$ term\cite{pertU}.
 
\paragraph{The model on a complete graph.}

 In this case, we can write the Hamiltonian (\ref{eq:Heff}) as
\begin{equation}
 {\cal H}_{\rm eff} = N_e\varepsilon_-
-\frac{t}{2S_c+1}\sum_\alpha F^\dagger_\alpha F^{\phantom{}}_\alpha 
\end{equation}
\[
      - \frac{2 t^2}{J_H(2S_c+1)^3} 
              \sum_{\alpha,\beta} 
              F^\dagger_\alpha \left[ 
                \left( N S_c + \hat N_e/2
                \right)\delta_{\alpha\beta}
              - {\bf S} \bbox{\sigma}_{\alpha\beta} 
              \right] F^{\phantom{}}_\beta ,
\]
where $\varepsilon_-=t- J_H S_c/2$, and we introduced the 
fermionic operators 
$F^\dagger_\alpha =\sum_{j} b^\dagger_{j\alpha} f^\dagger_j$ and 
$F_\alpha =\sum_{j} b_{j\alpha} f_j$. 
with the nonvanishing anticommutation relations 
\begin{equation}
  \left\{F^\dagger_\alpha, F^{\phantom{\dagger}}_\beta \right\} = 
  \hat Y \delta_{\alpha\beta}
  + {\bf S}\bbox{\sigma}_{\alpha\beta},
 \label{eq:ar_xed}
\end{equation}
where the spin operators ${\bf S} = \sum_j {\bf S}_j$ satisfy the usual 
\begin{equation}
  [S^+,S^-] = 2S^z, \quad [S^z,S^\pm] = \pm S^\pm, 
\end{equation}
su(2) spin algebra,
and with the $\hat Y$ operator defined as
\begin{equation}
  \hat Y = (S_c+1)N-\hat N_e/2 ,\qquad [{\bf S},\hat Y] = 0 .
\end{equation}
Finally, the commutation relations between the fermionic and bosonic operators read:
\begin{equation}
 [F^{\phantom{\dagger}}_\alpha,{\bf S}] = 
\frac{1}{2} \sum_\beta \bbox{\sigma}_{\alpha\beta} F^{\phantom{\dagger}}_\beta,
  \quad  
  [F^{\phantom{\dagger}}_\alpha,\hat Y] = -F^{\phantom{\dagger}}_\alpha/2, 
 \label{eq:cr_mixed}
\end{equation}
with their conjugate.
The set of relations (\ref{eq:ar_xed})-(\ref{eq:cr_mixed}) define a spl(2,1) 
graded algebra\cite{rittenberg76,Hubbardop}. 

 The rank of spl(2,1) is 2, 
and we can define two linearly independent Casimir 
operators\cite{rittenberg76}:
\begin{eqnarray}
  \hat C_2 &=& {\bf S}^2-\hat Y^2
   -\frac{1}{2} \sum_\alpha
    \left( F^\dagger_\alpha F^{\phantom{\dagger}}_\alpha 
         - F^{\phantom{\dagger}}_\alpha F^\dagger_\alpha
    \right) ,
  \nonumber
  \\
  \hat C_3 &=& 
   \left( 3 \hat C_2 \!-\! \hat S^2 \!+\! \hat Y^2 \!-\! \hat Y  \right)
   \left(\frac{\hat Y}{2} \!-\! \frac{1}{4} \right) 
   +  \frac{1}{2} \sum_{\alpha\beta} 
    F^\dagger_\alpha {\bf S}\bbox{\sigma}_{\alpha\beta} 
    F^{\phantom{\dagger}}_\beta .
  \nonumber
\end{eqnarray}
 In general case the irreducible representation $[Y,S]$ is 
$8S$ dimensional and contains the $(S,Y)$, $(S-1/2,Y+1/2)$, $(S-1/2,Y-1/2)$ and
$(S-1,Y)$ spin multiplets.
 Special cases concern the irreducible representation 
$[Y=S,S]$ which contains only the $(S,Y)$ and $(S-1/2,Y+1/2)$ spin
 multiplets (dimension  $4S+1$) and the irreducible representation
$[Y,S=1/2]$ which do not contain the $(S-1,Y)$ spin multiplet.
Applying operators $F_\alpha$ and $F^\dagger_\alpha$ we 
can walk between the different $(S,Y)$ spin multiplets within an 
irreducible representation. The eigenvalues of the Casimir 
operators $\hat C_2$ and $\hat C_3$ are $S^2-Y^2$ and $(S^2-Y^2)Y$ in the irrep $[Y,S]$, respectively, while the eigenvalues of operators $\hat {\bf S}^2$ 
and $\hat Y$ are $S'(S'+1)$ and $Y'$ for an $(S',Y')$ multiplet.

We can find the eigenvalues of the effective Hamiltonian using the 
representations of the spl(2,1) superalgebra.
The Hamiltonian conserves both the electron number and total spin:
 $[{\cal H},\hat Y]=[{\cal H},{\bf S}]=0$, however it does not commute with the $F$'s. 
On the other hand, since 
$
   \sum_\alpha F^\dagger_\alpha F^{\phantom{\dagger}}_\alpha 
  = \hat {\bf S }^2 - \hat Y^2 +\hat Y-\hat C_2,
\ $
the $(S\! -\! 1/2,Y\! -\! 1/2)$, $(S,Y)$, $(S\! -\! 1,Y)$, 
and $(S\! -\! 1/2,Y\! +\! 1/2)$ multiplets of the 
irreducible representation $[Y,S]$ are eigenstates of 
$\sum_\alpha F^\dagger_\alpha F^{\phantom{\dagger}}_\alpha$
with the eigenvalues
$2Y\! -\! 1$, $Y\! +\! S$, $Y\! -\! S$, and 0, respectively.
Similarly, the $\hat C_3$ appears in 
 the $t^2/J_H$ corrections in the Hamiltonian.
Finally, the effective Hamiltonian turns out to be a combination of 
$\hat C_2$, $\hat C_3$, $\hat {\bf S}^2$ and $\hat Y$, and the states 
$|(S,Y)_{[Y',S']}\rangle $ are eigenstates with (increasing) energies:
\beginwide
\begin{eqnarray}
 \label{eq:0nes}
  E_{[Y+1/2,S+1/2]}(S,Y) &=& N_e\varepsilon_- 
  -2 \left( t + \frac{2t^2}{J_H} \frac{S^{\rm max}+1}{(2S_c+1)^2}\right)
    \left( N-\frac{S^{\rm max}}{2S_c+1} \right)
 +\frac{4t^2}{J_H} \frac{S(S+1)}{(2S_c+1)^3}
  +{\cal O}(\frac{t^3}{J_H^2}),
 \\
  E_{[Y,S]}(S,Y) &=& N_e\varepsilon_- 
        - \left( t + \frac{2t^2}{J_H}\frac{S^{\rm max}-S}{(2S_c+1)^2} \right)
          \left( N - \frac{S^{\rm max}-S}{2S_c+1} \right) +{\cal O}(\frac{t^3}{J_H^2}),
 \\
  E_{[Y,S+1]}(S,Y) &=& N_e\varepsilon_- - 
     \left( t + \frac{2t^2}{J_H} \frac{S^{\rm max}+S+1}{(2S_c+1)^2} \right) 
 \left( N - \frac{S^{\rm max}+S+1}{2S_c+1} \right) +{\cal O}(\frac{t^3}{J_H^2}),
 \\
  E_{[Y-1/2,S+1/2]}(S,Y) &=& N_e\varepsilon_- ,
 \label{eq:enes}
\end{eqnarray}
\endwide
where $S^{\rm max}=N S_c + N_e/2=(2S_c+1)N - Y$ is the maximum 
total spin. 
In Fig.~\ref{fig:1} these energies are compared with those of the 
Kondo lattice for $U=J_H/2$ at left. It appear that the computed correction
is the correct one but for $t/J_H>0.05$ higher terms are necessary.
It is interesting to note that the
$E_{[Y-1/2,S+1/2]}(S,Y)= \varepsilon_- N_e$ is the eigenvalue for any $J_H$. 
To obtain this solution it is essential that the effective Hamiltonian 
can be 
expressed using the operators of the spl(2,1) superalgebra which, for
finite $J_H/t$, is possible for $U = J_H/2$ only. 

Let us now address the question of the degeneracy of each level.
For one site, the states with $S=S_c$ and 
$S_c+1/2$ form the irreducible representation 
$[S_c+1/2,S_c+1/2]$. With $N$ sites, the degeneracy 
$M^{(N)}_{[Y,S]}$ is determined by the number of times
the irrep $[Y,S]$ is 
contained in $[S_c+1/2,S_c+1/2]^N$. This is determined from the 
branching rule \cite{rittenberg76}, leading to the following recursion 
relation:
\[
 M^{(N+1)}_{[Y,S+1/2]} = \sum_{y=\{0,1/2\}}
   \sum_{s=-S_c-y}^{S-|S-y-S_c|} M^{(N)}_{[Y+y-S_c-1,S-s+1/2]} ,
\]
with $M^{(N)}_{[Y=N(S_c + 1/2),S]}=\delta_{S,N(S_c+1/2)}$ as boundary condition.
This recursion formula along with the energy definition in Eqs. (\ref{eq:0nes})-(\ref{eq:enes})
allows an iterative procedure to build the energy density distribution of the model. In Tab.~\ref{tab:1} we show as an example the spectrum of the 4 
site cluster.

For $N_e=1$ the $(S,Y)$ spin multiplet of the $[Y+1/2,S+1/2]$ is missing and 
the ground state is the highest spin state in the $[Y,S]$ irrep. 
For $1<N_e\leq N$ the $t/J_H$ correction makes
the lowest energy state to be the singlet in the $[Y+1/2,1/2]$ irrep, and the low
energy spectrum behaves as $S(S+1)$, like in the antiferromagnetic 
infinite range Heisenberg model.
For $N_e>N$ we can use that 
$E(N_e,t,J_H)=E(2N-N_e,-t,J_H)+U(N_e-N)$, and from this we obtain that when 
$N_e=N+1$ the ground state is again the highest spin state (a similar 
 situation is encountered in the Hubbard model\cite{nagaoka}). Finally,
for $N+1<N_e< 2N$ the ground state is highly degenerate  and the corresponding
energy can be obtained from Eq.~(\ref{eq:enes}) with the appropriate change.
To summarize, the highest spin multiplet is the non degenerate ground state
for $N_e=1$ and $N_e=N+1$ only, cases referred to true ferromagnetism.

 \paragraph{Semiclassical limit.}
In the limit $S_c \to \infty$ we can replace 
${\bf S}_j$ by $S_c (\sin \theta_j \cos \phi_j,\sin \theta_j \cos \phi_j,
\cos \theta_j)$ in the Kondo Hamiltonian (\ref{eq:HKLint}) with $J = J_H (S_c+1/2)$ fixed, and we end up 
with an noninteracting model\cite{classicalKL}. 
The one-particle spectrum comes out as follows: two $N-2$ degenerate states 
with energy $ \varepsilon_{\pm} = t \pm J/2$ and four nondegenerate 
states with energies 
\[
  \varepsilon_{s_1s_2} = -\frac{N-2}{2}t 
    + \frac{s_1}{2}\sqrt{J^2+N^2t^2-2 s_2 Jt \frac{S}{S_c}} \ ,
\]
where $s_1$ and $s_2$ are $\pm 1$.
The lowest energy state corresponds to $s_1=-1$ and $s_2=-1$, and 
for $J_H \gg t$ the energy is linerly decreasing with $S$, i.e. the fermions 
can freely propagate when the spins are parallel. In the 
next state ($\varepsilon_{-+}$)
the tendency is reversed: energy is higher for larger $S$. 
The absence of the true ferromagnetic state can be traced to the 
cancelation of the contributions linear in $S$ for $N_e>2$.
 This does not happen in a model with less pathological one-particle spectrum,
e.g  on the cubic lattice the ferromagnetism is realized for
a wide range of hole concentration\cite{dagotto}.
 Filling the one-particle levels, the lowest part
of the spectrum 
for $1<N_e\leq N$ electron is 
$\varepsilon_{--}+\varepsilon_{-+}+(N_e-2) \varepsilon_-$,   
$\varepsilon_{--}+(N_e-1) \varepsilon_-$,  
$\varepsilon_{-+}+(N_e-1) \varepsilon_-$, and $N_e \varepsilon_-$.
These energies are equal, up to corrections ${\cal O}(t^3/J^2,1/S_c)$, to the energies 
(\ref{eq:0nes})-(\ref{eq:enes}) of the $(S,Y)$ spin multiplet in the 
$[Y+1/2,S+1/2]$, $[Y,S]$, $[Y,S+1]$, and $[Y-1/2,S+1/2]$ 
irreps, respectively. This way we can establish a one-to one 
correspondence between the semiclassical and quantum spectra for this
 particular model.

 To conclude, we have shown that the effective strong coupling limit of the
Kondo lattice model on a complete graph can be described by the spl(2,1) 
graded algebra and exhibits a dynamical supersymmetry.
In contrast to Hubbard model, the supersymmetry is not limited to the
$t$-model only, but holds also 
in the next order ($t^2/J_H$) for a special value of on-site repulsion. 
This, and the numerical diagonalization on small clusters suggest that not 
only the strong coupling limit, but the Kondo Hamiltonian itself is 
supersymmetric.
 We also show that for $S_c \to \infty$ the spectrum of quantum model can 
be derived from one-particle states of the semiclassical Kondo lattice
model. Finally, the ferromagnetic ground state is reduced to the 
case when we have only 1 or N+1 electron, in contrary to the generally 
accepted 
expectations. 

 We would like to acknowledge useful discussions with  P. Fazekas, 
K. Hallberg, K. Kubo, F. Mila, E. M\"uller-Hartmann, H. Shiba, and R. Shiina.

\begin{table}
\begin{tabular}{l|ccc|ccc|cccc|cccc|ccccc}
$N_e(Y)$  
     &  \multicolumn{3}{c|}{0(6)} 
     &  \multicolumn{3}{c|}{1(11/2)} 
     &  \multicolumn{4}{c|}{2(5)} 
     &  \multicolumn{4}{c|}{3(9/2)}
     &  \multicolumn{5}{c}{4(4)} \\
$E'~~~ 2S$& 4&2&0& 5&3&1&  6 & 4 & 2 & 0 & 7 &5 &3 &1 & 8&6&4&2&0\\
\hline
    0& 1&3&2&  3 & \underline{8} & 7 &  3 & 6 & 6 & 3 &  1 &   &   &   &  & & & & \\
 -1/2&  & & &    &   &   &    &   &   &   &    & 3 &   &   &  & & & & \\
   -1&  & & &    &   &   &    & 3 &   &   &    &   & 6 &   &  & & & & \\
 -3/2&  & & &    & 1 &   &    &   & \underline{8} &   &    &   &   & 6 &  & & & & \\
   -2&  & & &    &   & 3 &    &   &   & 7 &    &   &   &   &  & & & & \\
 -5/2&  & & &    &   &   &    &   &   &   &    &   &   & 3 &  & & & & \\
   -3&  & & &    &   & 2 &    &   & 7 &   &    &   & 6 &   &  & & & & \\
 -7/2&  & & &    & 3 &   &    & \underline{8} &   &   &    & 6 &   &   &  & & & & \\
   -4&  & & &  1 &   &   &  3 &   &   &   &  3 &   &   &   & 1&3&6&6&3\\
 -9/2&  & & &    &   &   &    &   &   &   &    & 3 & \underline{8} & 7 &  & & & & \\
   -5&  & & &    &   &   &    & 1 & 3 & 2 &    &   &   &   &  & & & & \\
\end{tabular}
\caption{The spectrum and multiplicities of the quantum double exchange ($J_H/t \to + \infty$) for
 $S_c =1/2$ and $N=4$. Here $E'=E-N_e\varepsilon_-$. 
As an example, we have underlined the 
 four spin multiplets of the $[Y=5,S=2]$ irreducible representation, which has 
multiplicity 8.}
\label{tab:1}
\end{table}

\end{multicols}

\begin{references}
\bibitem{Hubbardop}
 When $S_c=0$, the $F$ operators can be identified with the Hubbard 
operators as $F^\dagger_\alpha = X_{\alpha 0}$ and 
$F^{\phantom{\dagger}}_\alpha = X_{0\alpha}$, 
and this provides a direct link to the solution of the $t$-model as 
presented by Kirson in \cite{kirson}.
\end{references}
\end{document}